\documentclass[12pt,twocolumn,a4paper]{article}
\usepackage[T1]{fontenc}
\usepackage[utf8]{inputenc}
\usepackage{mathptmx}
\usepackage[a4paper,left=41pt,right=41pt,top=48pt,bottom=56pt,headsep=14pt,headheight=14pt,columnsep=15pt]{geometry}
\usepackage{amsmath,graphicx,booktabs,fancyhdr,microtype}
\usepackage[labelfont=bf,labelsep=period,font=small]{caption}
\usepackage{titlesec,url,xcolor,flushend}
\usepackage[colorlinks=true,urlcolor=blue!50!black,linkcolor=black,citecolor=black,pdftitle={Electric Propulsion Database Development and Integrated Modelling Analysis},pdfauthor={Talha Baig}]{hyperref}
\renewcommand{\thesection}{\arabic{section}.}
\renewcommand{\thesubsection}{\arabic{section}.\arabic{subsection}.}
\titleformat{\section}{\large\bfseries}{\thesection}{.5em}{}
\titleformat{\subsection}{\normalsize\bfseries}{\thesubsection}{.5em}{}
\titlespacing*{\section}{0pt}{8pt}{3pt}
\titlespacing*{\subsection}{0pt}{6pt}{2pt}

\begin{document}
\twocolumn[{
\begin{center}
{\LARGE\bfseries Electric Propulsion Database Development and Integrated Modelling Analysis\par}
\vspace{6pt}
{\normalsize Talha Baig$^a$, Dr Kate Smith, Dr Nicholas Crisp\par}
{\small $^a$Department of Mechanical, Aerospace and Civil Engineering, The University of Manchester, Manchester M13 9PL, United Kingdom\par}
{\small\href{mailto:mirza.baig-7@postgrad.manchester.ac.uk}{mirza.baig-7@postgrad.manchester.ac.uk}\par}
\end{center}
\vspace{4pt}
\noindent \textbf{Abstract.} Electric propulsion (EP) systems require significant subsystem integration and data driven modelling for accurate sizing. This paper is an overview of a high-fidelity database development and analysis aiming to fill the gap in electric propulsion sizing. This is a preliminary study to develop a tool that considers all thruster technologies that exist within Electric Propulsion. Since thrust is set by the product of efficiency and available electrical power, this paper showcases that thruster selection directly constrains the spacecraft power bus. We showcase and hypothesize within the database analysis and in the future work that the correct way to choose electric propulsion systems is through budgeting the EP system through Mission constraints. Comparing technologies for propulsion parameters is important for spacecraft optimisation and no such dataset exists. The literature offers either deep surveys of one thruster type, which cannot be compared across technologies, or broad catalogues that quote manufacturer figures at a single operating condition without checking whether those figures agree with one another. This paper reports the development, performance levels and preliminary analysis of a unified Electric Propulsion technology database. Every device is reduced to a common set of parameters including power, thrust, specific impulse, exhaust velocity and total efficiency linked by governing analytical equations and state space models. The database holds 390 distinct models from 22 countries across twelve thruster types, from 0.4 W to 1 MW, with first flights from 1964 to 2025. The database was run through a consistency check which suggested that 180 of 433 disagreed with the analytical relation. Then parametric relations were developed through a logarithmic function coupled with statistical processing and Gaussian uncertainty analysis. This provided significant islands across analytical relations that were bounded by the quantity of datasets from the thruster database.

\vspace{4pt}

\vspace{4pt}\hrule\vspace{8pt}
}]
\thispagestyle{fancy}
\section{Introduction}

\noindent Electric propulsion (EP) systems convert electrical energy into propellant kinetic energy through electrothermal, electrostatic, or electromagnetic acceleration. (Krejci and Lozano 2018) Relative to chemical propulsion, EP can provide a substantially higher specific impulse and reduce propellant mass. (O’Reilly et al. 2021).

However, EP systems generally deliver low thrust for the electrical power available. Technology selection must therefore consider power, thrust, specific impulse, efficiency, mass, lifetime, maturity and evidence quality together (Goebel and Katz 2008).

With electric propulsion sizing and selection there is a significant gap in the maturity of technology rather than the data we have. The lack of maturity creates major issues in verifying and proving the accuracy of the data sets provided by the manufacturers. Hence, there is limited information and analysis for understanding how overstated or understated database numbers are against analytical predictions. If a review may quote a representative technology range, a manufacturer may publish a nominal product value, a test paper may report a throttle map with uncertainty and facility conditions. Combining these sources without retaining their provenance can create false precision, unmatched operating points and maturity claims that are falsely stated.

Database design is the preliminary task for the overarching scope of a high-fidelity tool development. Hence developing a tailored and high fidelity database is a significant part of the research problem. As the database development process and a comparative set of thrusters across all EP types would allow to quantitatively visualise and map the gaps and relations with technology maturity and EP performance. It is really important to attribute and analyze the characteristics of this database to keep it reproducible and useful for the toolkit. The database has citations attached to a row which establishes that a document has been identified, but not that every stored field came from that document, that the value was read rather than inferred, or that power, thrust and specific impulse refer to the same operating point. Likewise, a technology readiness level (TRL) is not a generic property of a device name: it depends on a defined system, application, environment and body of evidence (National Aeronautics and Space Administration 2020). These distinctions must remain visible if the data are to support scholarly analysis or a traceable engineering tool.

This paper develops and evaluates a multiple thruster comparative layer for EP with aims to be used by mission designers for integrating the most optimum thruster strings for their mission designs. There have been multiple additional databases developed and there are a handful of integrated propulsion studies that exist. However this review and database analysis extends beyind literature and database studies to extend to all thruster types, cover a wide variety of missions and spacecraft sizes and utilise that broad scope with an aim to develop a high fidelty multi modal integrated EP sizing tool with all EP families in the tool box. Broad surveys, thruster-specific databases and measurement archives are established foundations. Therefore, the primary gap becomes narrower where few to none public resources combine broad cross thruster type coverage with explicit evidence, reproducible transformations and tests showing whether engineering conclusions change when weaker evidence is removed.

The contribution has four parts. First, the work distinguishes the 433 operating-point rows inherited from the legacy compilation from the 390 named systems in the final release. Second, every released record is tied to a full citation. Third, all data transformations, classification rules and physics-based consistency checks are implemented in a reproducible processing pipeline. Each physical relation is applied only where its assumptions are valid, and statistical resampling uses fixed random seeds so that the reported results can be regenerated. Fourth, sensitivity to source quality is assessed by comparing the complete dataset with records verified against their cited documents. In the thrust–power regressions, records that cite the same document are resampled as a group because they may not be statistically independent. The principal contribution is therefore a traceable link between source evidence, database construction and engineering analysis.

The study asks: (i) what technologies and performance ranges are represented; (ii) how traceable and reproducible the release is; (iii) what are the challenges with source quality and does it impact scaling and integrated modelling; and (iv) which engineering uses the present schema can support safely. Counts describe this compiled convenience sample, not the global EP population or market share. Each row represents a named model or system, not a complete throttle map. The paper therefore focuses on sensitivity and data limitations to give an overview of the various technologies developed within each thruster type. This would allow to answer the questions around which technologies are holistically more mature and why and which thruster types have what specific technologies that have a high TRL and why?

\section{Related resources and research gap}

\subsection{Broad surveys}

\noindent Several publications review propulsion technologies for small spacecraft, although their aims and coverage differ. Krejci and Lozano (2018) review the working principles, spacecraft constraints and miniaturisation potential of propulsion concepts for small spacecraft. Levchenko et al. (2018) examine micropropulsion for CubeSats and small satellites, with particular attention to miniaturised plasma and electric thrusters, their physical principles and their testing requirements. O’Reilly et al. (2021) compare electrothermal, electrostatic and electromagnetic electric-propulsion methods for small-satellite missions, including their principal performance characteristics and applications.

The 2026 edition of NASA’s\textit{ }\textit{State-of-the-Art}\textit{ }\textit{of} \textit{Small Spacecraft Technology} provides a broader catalogue of in-space propulsion devices. It summarises reported performance, development status and flight history using publicly available information (NASA Small Spacecraft Systems Virtual Institute 2026). The report states that much of this information originates from manufacturer literature and that independent verification of device performance should not be assumed unless it is explicitly reported (NASA Small Spacecraft Systems Virtual Institute 2026). These limitations support the need for clear source traceability and evidence-based maturity assessment.

The principal strength of broad surveys is their technology coverage. They help readers identify available propulsion concepts, understand their possible mission applications and perform preliminary technology screening. However, they generally report representative product parameters rather than measurements obtained under consistent test and throttle conditions. They should therefore not be treated as controlled experimental datasets without further checks on parameter definitions, source quality and test conditions.

The present database is also derived from published literature and publicly available technical information; it is not a standardised experimental campaign. Each record contains a representative set of thruster parameters.

Within the resources reviewed here, the remaining gap is the need for a reproducible, cross-technology EP dataset that combines a common parameter map with explicit source provenance, physical consistency checks and evidence assessment. The present study addresses this gap while retaining the limitations associated with literature-derived data.

\subsection{Physical lessons from technology-specific datasets}

\noindent Technology-specific datasets are valuable because they retain the variables that govern thruster performance. For a steady thruster at a matched operating point, and using consistent definitions of input power and total propellant mass flow,

\begin{equation}T=\dot{m}g_0I_{\mathrm{sp}},\qquad \eta_{\mathrm{T}}=\frac{Tg_0I_{\mathrm{sp}}}{2P_{\mathrm{in}}},\qquad T=\frac{2\eta_{\mathrm{T}}P_{\mathrm{in}}}{g_0I_{\mathrm{sp}}}.\end{equation}
(1) These relations show that power, thrust and specific impulse cannot be interpreted independently. Their relationship depends on the efficiency with which electrical power is converted into directed exhaust kinetic energy. This efficiency is influenced by the acceleration mechanism, propellant, geometry, magnetic-field topology, operating condition and loss processes. Equation 1 is valid only when the reported quantities refer to the same operating point and use compatible system boundaries. Mixing discharge power with total-system efficiency, or anode mass flow with total specific impulse, can produce physically inconsistent comparisons.

Hall-thruster scaling studies demonstrate the importance of retaining this physical context. Dannenmayer and Mazouffre (2009) compiled 33 single-stage Hall thrusters between 10 W and 50 kW and used their geometry and performance to determine scaling coefficients. The study showed that preliminary sizing requires discharge-channel dimensions, magnetic-field strength and assumptions concerning plasma density and wall temperature; input power alone is insufficient. Giammarinaro et al. (2023) subsequently developed a scaling method specifically for high-power magnetically shielded Hall thrusters. This indicates that even within one propulsion type, a change in magnetic topology can create significant differences in thruster physics and performance.

Data-driven Hall-thruster studies provide another lesson where prediction quality is controlled by the distribution of the available data. Hence, the larger the dataset and the more distributed the data, the more accurate the model. This will be explained in detail through Figure 1. Marianacci and Mazouffre (2024) assembled 54 stationary-plasma thrusters and 3323 operating conditions covering approximately 50 W to 70 kW. Their supervised-learning method performed well in data-rich regions but produced substantially poorer predictions in the sparsely sampled high-power region. Masillo et al. (2025) examined approximately 60 xenon Hall-effect-based designs below 1 kW, including annular, cylindrical, cusped-field and magnetically shielded configurations. The observed dispersion reflects both the difficulty of scaling Hall discharges to low power and the fact that these configurations do not share identical wall losses, ionisation behaviour or magnetic confinement.

Other propulsion types require different explanatory variables. Princeton’s applied-field MPD database retains thrust together with current, voltage and applied magnetic field because MPD performance depends directly on the discharge-current and magnetic-field regime (Princeton University Electric Propulsion and Plasma Dynamics Laboratory 2026). Pulsed-plasma-thruster comparisons require pulse energy, impulse bit, mass bit and electrode configuration; average power alone does not describe the transient acceleration process (Molina-Cabrera et al. 2011). Plume measurements introduce another layer of variation because the reported result depends on facility pressure, diagnostic position, probe design and data reduction (Reissner et al. 2013). Recommended thrust-measurement practice similarly shows that calibration, drift correction, uncertainty estimation and processing choices are part of the reported thrust value (Polk et al. 2017).

The resulting scatter in an EP tradespace therefore has several possible origins. It may represent real differences between acceleration mechanisms, variation between designs within one technology type, operation at different throttle points, facility and diagnostic effects, or differences in source quality. A cross-technology database can expose where this scatter occurs, but identifying its cause requires the relevant physical and evidential variables to be retained.

\subsection{Significance of data reproducibility in uncertainty analysis}

\noindent The use of provenance in this study allows an apparent physical trend to be tested against the quality and origin of its supporting evidence. The FAIR principles motivate persistent identifiers, structured metadata and reusable research objects (Wilkinson et al. 2016). The corresponding principles for research software extend these expectations to the scripts and workflows that generate the results (Barker et al. 2022). Reproducible-computation guidance further motivates preserved raw inputs, versioned transformations and data retained behind each figure (Sandve et al. 2013).

These practices allow the tradespace and uncertainty analysis to be repeated using different evidence subsets. For example, results obtained from the full database can be compared with results restricted to directly inspected primary sources. If a high-variance region disappears after manufacturer-only or historical records are removed, the variation is more likely to be evidence-dependent than a robust physical feature. The current release supports record-level source sensitivity, but it does not yet contain complete field-level source locators, reported measurement uncertainties or facility metadata for every model. These remain requirements for attributing observed variation to specific experimental causes.

\subsection{Database scope and physical variables}

\noindent The released database contains 390 named EP models. The main analysis therefore focuses on the physical and evidential properties of the database. Each model is described by its propulsion type and subtype, propellant, operating mode, power, thrust, specific impulse and efficiency where available. The database also retains reported performance ranges, selected electrical and lifetime variables, maturity and flight evidence, source category and bibliographic provenance. Missing quantities remain null. Where several operating points were available for one model, the number of points and the observed performance envelope were retained alongside a representative matched point.

This model-level structure supports comparison across the EP landscape, but it does not reproduce a complete throttle map. Consequently, the database is used for technology screening, coverage analysis and uncertainty mapping. It is not presented as a substitute for operating-point-resolved datasets used for detailed thruster design or plasma-model calibration.

\subsection{Physics-informed Gaussian-process analysis}

\noindent Only records for which power, thrust and specific impulse refer to a common operating point are used in the physics-informed analysis. An effective thrust efficiency is calculated as

\begin{equation}\eta_{\mathrm{eff},i}=\frac{T_ig_0\,{I_{sp}}_i}{2P_i}.\end{equation}
\noindent Values outside the physical interval 0$<$$\eta$\textsubscript{eff} $<$ 1 are returned for source review rather than clipped or silently included. Such values may indicate incompatible power definitions, unmatched throttle points, transcription errors or inconsistent mass-flow boundaries.

Gaussian-process regression is then applied to the residual structure in

\begin{equation}\log T_i=\log\left(\frac{2P_i}{g_0\,{I_{sp}}_i}\right)+f_k(\mathbf{x}_i)+\varepsilon_i,\end{equation}
\noindent where\textit{ }\textit{f}\textsubscript{\textit{k}} represents the efficiency-related residual for propulsion type\textit{ k}, and\textbf{ x}\textsubscript{\textit{i}} contains the available explanatory variables, such as power, propellant, subtype, maturity and evidence class. Because different EP mechanisms do not form a single smooth physical population, separate type-level models or a multi-task Gaussian process are used instead of one stationary kernel across all technologies.

The posterior variance of\textit{ }\textit{f}\textsubscript{\textit{k}} is interpreted as coverage-dependent model uncertainty under the adopted inputs and kernel. It is expected to increase in sparsely sampled or extrapolative regions. Input-dependent residual variance is modelled separately where the data support a heteroscedastic likelihood. This residual variation may contain genuine design variability, operating-point differences and unrecorded measurement effects; it is therefore not labelled as experimental uncertainty.

Model performance is assessed using blocked cross-validation across power regions and propulsion subtypes. Predictive intervals are evaluated for empirical coverage rather than reported without calibration. The analysis is repeated for the full and verified-source subsets to determine whether uncertainty patterns are robust to evidence quality.

\section{Database construction and methodology}

\subsection{Data lineage and unit of record}

\begin{table*}[t]
\caption{Physical and statistical lessons obtained from existing data-centric EP studies and their implications for the present database.}
\small\centering
\begin{tabular}{@{}p{0.19\textwidth}p{0.34\textwidth}p{0.41\textwidth}@{}}
\toprule
Study or resource & Main learning point & Implication for this study\\
\midrule
Dannenmayer and Mazouffre & Hall-thruster scaling depends on geometry, magnetic field and physical similarity assumptions, not power alone. & Cross-technology trends must retain the propulsion type and must not be interpreted as universal scaling laws.\\
Giammarinaro et al. & Magnetically shielded high-power Hall thrusters require a topology-specific scaling treatment. & Subtype and magnetic configuration can explain part of the variation within the Hall-thruster population.\\
Marianacci and Mazouffre & Machine-learning prediction deteriorates where operating-point coverage is sparse or unbalanced. & Predictive uncertainty should be mapped against data density and power range rather than reported as a single global error.\\
Masillo et al. & Low-power Hall-based designs display substantial dispersion because different configurations and loss mechanisms are combined. & Input-dependent scatter should be modelled, and distinct Hall configurations should not automatically share one regression function.\\
Princeton AF-MPD database & Current, voltage and applied magnetic field are needed to interpret MPD performance. & A model based only on power, thrust and $I_{\mathrm{sp}}$ can locate trends but cannot identify MPD acceleration physics.\\
European plume database and thrust-measurement practice & Facility conditions, calibration and diagnostic processing contribute to differences between reported values. & Unexplained scatter must not be labelled as intrinsic technology uncertainty without measurement and facility metadata.\\
Hall-thruster UQ studies & Bayesian calibration and sensitivity analysis can identify influential model inputs when the corresponding physical variables are available. & Source category, maturity and available physical metadata can be tested as explanatory variables, while unavailable causes must remain unresolved.\\
Present database & Broad coverage makes sparsely populated regions and evidence imbalances visible across EP technology types. & The database supports landscape-level uncertainty mapping, evidence-sensitivity analysis and prioritisation of future data collection, rather than detailed prediction of an individual throttle map.\\
\bottomrule
\end{tabular}
\end{table*}
\begin{table*}[t]
\caption{Development of the final 390-model database.}
\small\centering
\begin{tabular}{@{}p{0.2\textwidth}cp{0.6\textwidth}@{}}
\toprule
Step & Number & Explanation\\
\midrule
Initial source data & 433 records & Performance records describing 308 named systems. Some systems appeared in more than one record.\\
Initial EP set & 306 models & Two non-propulsive systems were removed, and repeated records for the same model were combined.\\
Additional models & 84 models & Models with a retrievable source were added. Six candidates without a confirmed source were excluded.\\
Final database & 390 models & The final database contains 306 initial models and 84 additional models.\\
\bottomrule
\end{tabular}
\end{table*}

\noindent Table 2 separates the successive data populations. The database was first compiled with 433 thrusters and then later on condensed to 390 models after further additions and reductions on data that was superficial or without explicit reference data. Two entries were removed because they were not propellant-expelling thrusters: a magnetorquer and an electrostatic plasma-brake concept. The remaining rows were collapsed to 306 named model records. Where a model had several source rows, a labelled nominal point was preferred; otherwise a physically reconcilable point nearest the within-model median power was retained. The number of operating points and the power, thrust and specific-impulse envelopes were stored so that the collapse remained visible.

90 additional thrusters were then added covering electrothermal, pulsed, electrostatic micropropulsion, Hall, MPD and recent commercial systems.

The released row comprises the following identifiers 1. model name 2. manufacturer; country 3. Thruster subtype 4. propellant 5. operation mode 6. nominal power, thrust, specific impulse and efficiency. 7. selected electrical, mass, lifetime and flight fields; performance envelopes 8. readiness fields 9. citation 10. URL or digital object identifier 11. source grade and 12. Additional notes

Electric Propulsion systems were graded in the form of 5 sources.\textit{ Primary} denotes a checked journal, conference or agency source;\textit{ datasheet plus pri-} \textit{mary} denotes a manufacturer value corroborated by a paper. Together these form the\textit{ verified} analysis class. \textit{Datasheet} becomes the\textit{ manufacturer} class;\textit{ compila-} \textit{tion} denotes a value reconciled from the cited source and adjacent literature; and\textit{ historical} denotes a real citation that was not directly viewed. These labels describe source handling, not experimental uncertainty or truth.

Traceability is assessed at four levels: a record can be named, cited, linked and claim-located. All released records are named and cited, and most are linked. The current database does not systematically connect each field to an exact page, table, figure or data row. Accordingly, the release has complete citations at row level.

\subsection{Operation mode and physical consistency}

\noindent The release distinguishes 355 steady, 28 pulsed and seven duty-cycled systems; it also distinguishes 380 electrically powered from ten chemically augmented records. These categories govern whether steady electrical identities may be applied. For matched steady operation,

\begin{equation}T=\dot{m}g_0I_{\mathrm{sp}},\end{equation}
\noindent where\textit{ }\textit{T} is thrust, ˙\textit{m} is propellant mass flow and \textit{g}\textsubscript{0} $=$ 9$.$80665 m s$^{-}$\textsuperscript{2}. With total jet efficiency $\eta$ and electrical input power\textit{ P},

\begin{equation}\eta^*=\frac{Tg_0I_{\mathrm{sp}}}{2P},\qquad\frac{T}{P}=\frac{2\eta}{g_0I_{\mathrm{sp}}}.\end{equation}
Equation 5 is used as a consistency rule only for steady electric records with positive, matched fields. If a stored efficiency differed from $\eta$$^{*}$ by more than 5\% while $\eta$$^{*}$ $\leq$ 0$.$90, the released value was recomputed and the action recorded. A value above 0.90 was marked irreconcilable and set to missing rather than forced into range. Expansion-record efficiencies were derived only where the relation applied. At model level, 156 values are labelled agreed, 129 recomputed, 59 derived, 35 not applicable, seven chemically augmented and four irreconcilable.

This processing is intentionally visible because it limits interpretation. The 341 efficiency values used in Figure 3 for steady electric systems are predominantly algebraic descendants of the same\textit{ P},\textit{ T} and\textit{ I}\textsubscript{sp} fields. Their agreement with Equation 5 verifies transformation consistency; it is not independent experimental validation.

\subsection{Technology Readiness Levels (TRLs)}

\noindent The database assigns a rule-based TRL from the documented development status of each model. A quantified concept with no reported hardware test is assigned TRL 3. Ground testing at one documented condition gives TRL 4, while testing over several throttle conditions gives TRL 5. An in-space demonstration is assigned TRL 7, qualified hardware TRL 8, and hardware with documented operational flight use and a flight year TRL 9. Ground-tested models are capped at TRL 5 because the available fields do not show whether a prototype was demonstrated in a relevant environment. Each assignment retains a short evidence statement and confidence grade.

These values provide a consistent maturity classification within the database where they are not formal programme-level assessments. A formal assessment requires a defined system, intended application, relevant environment and reviewed evidence package (National Aeronautics and Space Administration 2020). TRL and first-flight year are analysed separately because maturity and documented flight are different properties. The grouped TRL 7–9 category includes demonstrations, qualified hardware and flight-proven systems, so it does not by itself confirm successful propulsion operation. TRL is treated as an ordered category and is not averaged.

\subsection{Statistical analysis}

\noindent The statistical analysis is used to describe the database, compare thrust–power trends between thruster types and test how strongly those trends depend on the included records and sources. Counts use all relevant records; medians and ranges use available non-missing values. Logarithmic axes are used because power and thrust span several orders of magnitude.

For each steady, electrically powered thruster type with at least eight eligible models, nominal thrust is fitted against nominal input power as

\begin{equation}T=aP^b,\end{equation}
\noindent where\textit{ }\textit{a} sets the position of the curve and\textit{ }\textit{b} is the scaling exponent. Taking logarithms gives log \textit{T} $=$ log \textit{a} $+$ \textit{b} log \textit{P}. Ordinary least squares then finds the straight line that minimises the squared differences between recorded and fitted log \textit{T}. An exponent of \textit{b} $=$ 1 indicates proportional thrust–power scaling; \textit{b} $>$ 1 and\textit{ b} $<$ 1 indicate faster and slower growth, respectively. This is a comparison between different models, not a throttle law for one thruster.

The fit is summarised by\textit{ R}\textsuperscript{2} and the scatter factor \textit{e}$^{\sigma}$. The value of\textit{ R}\textsuperscript{2} measures how closely the logarithmic data follow the fitted line, while\textit{ e}$^{\sigma}$ gives the multiplicative spread of models around it. In Figure 2, solid lines use all eligible records and dashed lines use the TRL 7–9 subset where at least five models are available. The shaded region shows the residual spread ($\pm$1$.$96$\sigma$ in logarithmic space); it is not a confidence band.

Two resampling checks are used to test how stable the exponent\textit{ b} is. Both work the same simple way: redraw a new sample from the available records, allowing a record to be picked more than once, refit the line, and repeat many times. The spread of exponents this produces shows how much the result would change under a slightly different sample. A narrow spread means a stable result; a wide spread means the result is sensitive to exactly which records are included.

The first check, used in Figure 2, resamples individual records directly: 4,000 new samples are drawn and refitted, giving 4,000 estimates of\textit{ b}, and the middle 95\% of these forms the reported interval. The second check resamples whole documents instead of individual records, because one publication can support several models and those models are then not independent pieces of evidence. Records that cite the same document are always kept or dropped together. This document-level check is repeated 3,000 times and compares the complete dataset with the verified-source subset where enough models are available, showing whether the fitted exponent depends on a small number of publications or on less-verified sources.

These intervals show how strongly the fitted exponent depends on the compiled records and cited sources. They do not measure experimental uncertainty or define a universal propulsion law. Formal hypothesis tests are not used because the database is a literature compilation, not a random sample or controlled experiment. The regressions are therefore used only for comparison and preliminary screening within the observed power range of each thruster type.

\subsection{Reproducible build and validation}

\begin{table}[t]
\caption{Released records by thruster type. “Verified” is a source-use class; “TRL 7–9” is the database readiness band.}
\small\centering\setlength{\tabcolsep}{4pt}
\begin{tabular}{@{}lrrrr@{}}
\toprule
Family & $n$ & Verified & TRL 7–9 & Linked\\
\midrule
Hall & 203 & 122 & 108 & 189\\
Gridded ion & 65 & 52 & 50 & 56\\
MPD & 23 & 11 & 1 & 20\\
Pulsed plasma & 22 & 15 & 16 & 8\\
FEEP & 18 & 14 & 14 & 11\\
Helicon & 13 & 5 & 8 & 13\\
Arcjet & 11 & 10 & 6 & 2\\
Resistojet & 10 & 8 & 9 & 1\\
Other electrothermal & 9 & 1 & 9 & 8\\
Vacuum arc & 8 & 3 & 7 & 5\\
Electrospray & 5 & 2 & 2 & 0\\
ECR & 3 & 3 & 0 & 0\\
\midrule
All released & 390 & 246 & 230 & 313\\
\bottomrule
\end{tabular}
\end{table}

\noindent The database is produced by an ordered pipeline that reads the preserved source files, standardises fields and units, applies the selection and classification rules, connects records to references, writes intermediate files and creates the final database, figures and tables. The process is deterministic: the same inputs, software environment and settings produce the same outputs.

\section{Results}

\subsection{Coverage, completeness and traceability}

\noindent The release contains 390 unique names in 12 technology classes, from 161 manufacturers, 22 countries and 23 propellants. It uses 293 distinct citation texts. Every row has a citation, while 313 (80.3\%) also include a direct link or digital object identifier. Hall thrusters account for 203 records (52.1\%) and gridded-ion thrusters for 65 (16.7\%); together they form 68.7\% of the release. Cross-family summaries must therefore expose rather than conceal this imbalance.

Core numerical fields are complete: all 390 rows contain nominal power, thrust, specific impulse and rule-assigned TRL. Efficiency is present for 358 rows (91.8\%), dry mass for 377 (96.7\%) and first-flight year for 134 (34.4\%). The evidence classes comprise 246 verified, 49 manufacturer, 75 compiled and 20 unviewed records. Coverage is consequently broad, but neither uniform nor equally traceable across families.

\subsection{Multiple Thruster Performance Comparative Analysis}

\noindent Recorded power spans 0.4 W to 1 MW, with a median of 715 W. Thrust spans 5$\times$ 10$^{-}$\textsuperscript{4} mN to 10\textsuperscript{5} mN, with a median of 38 mN; specific impulse spans 48–14,000 s, with a median of 1,700 s. These ranges exceed those of a typical product survey because the release includes concepts, laboratory hardware, microthrusters and high-power studies.

Figure 1 shows recognisable technology regimes. Electrothermal devices occupy comparatively low specific impulse; FEEP and electrospray devices extend towards low thrust; and the highest-power MPD records reach the upper-right region. The dense Hall and gridded-ion bands dominate the central sample. These patterns justify family-aware screening, but not a pooled ranking: much of the apparent separation is created by acceleration physics, power range and collection coverage together.

\begin{figure*}[t]
\centering
\includegraphics[width=\textwidth]{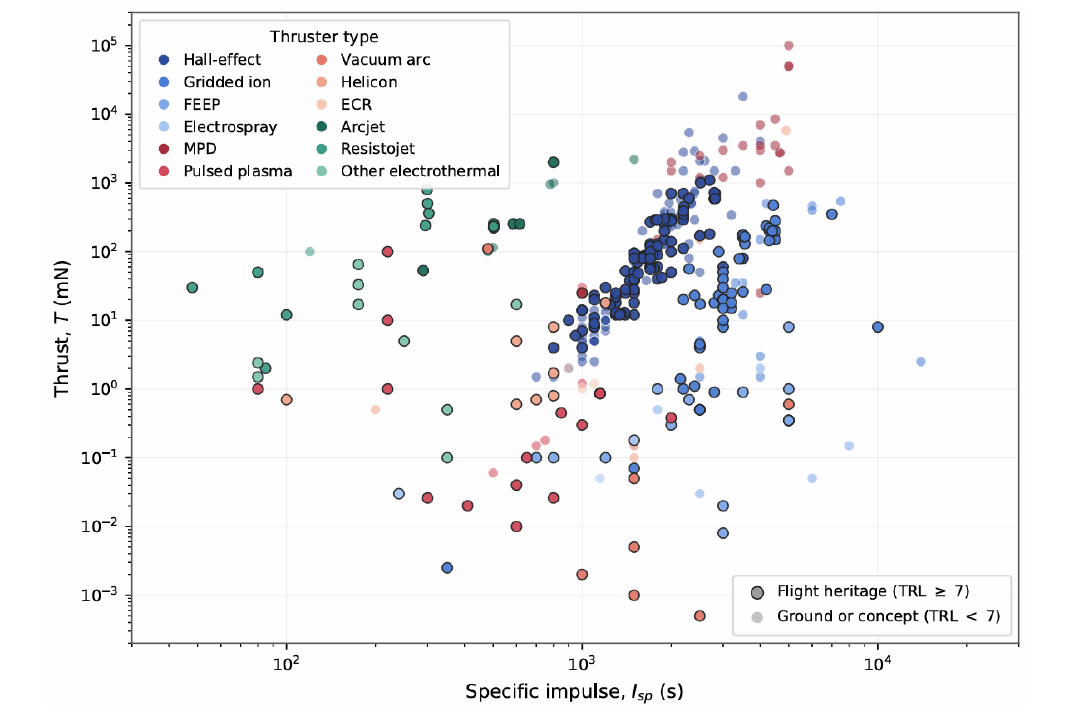}
\caption{Thrust–specific-impulse overview of the database ($n=390$). The figure showcases all 390 thrusters color coded based on the thruster type. Solid spots are at or above TRL 7. Slightly smaller and translucent spots are lower than TRL 7.}
\end{figure*}

The plot provides an empirical search envelope for a future sizing workflow. It does not by itself identify an optimum technology, because it excludes lifetime, total impulse, power-processing mass, plume constraints, propellant storage, thermal rejection, cost and qualification evidence. Interpolation should be restricted to a relevant family and power regime.

\subsection{Thrust-to-power ratio and efficiency}

\noindent Equation 5 makes the central EP trade explicit: at fixed total efficiency, thrust per unit electrical power decreases inversely with specific impulse. Hall thrusters occupy a comparatively compact band, while gridded-ion systems extend to higher specific impulse and lower thrust-to-power ratio. Electrothermal devices occupy the high-thrust-to-power, low-specific-impulse region.

\subsection{Thrust–power scaling and evidence sensitivity}

\begin{figure*}[t]
\centering
\includegraphics[width=\textwidth]{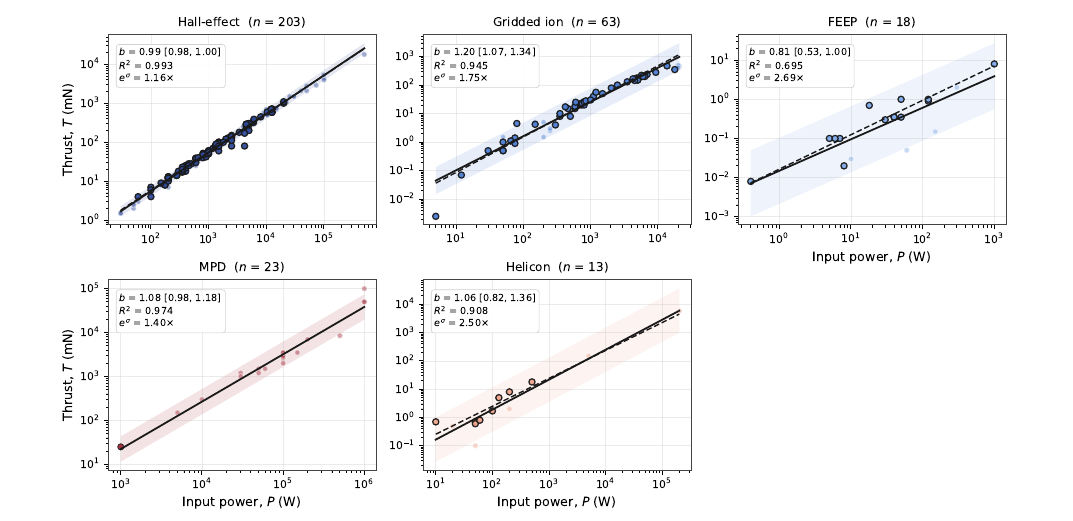}
\caption{Thrust–power fits, $T=aP^b$, for the steady electric families with at least eight eligible records. For each family, a straight line is fitted to $\log T$ against $\log P$ by ordinary least squares; $a$ and $b$ are recovered from that line as described in Section 3.4. Solid curves use all eligible records, and the shaded band around each solid curve shows the typical scatter of the underlying points ($\pm1.96$ standard deviations in log space, not a confidence interval). Dashed curves repeat the fit using only the higher-readiness (TRL 7–9) subset, where at least five models are available. The bracketed values give the 95\% interval from resampling individual records 4,000 times, which shows how much $b$ would change under a slightly different sample of records. These are descriptive fits to the compiled sample, not universal physical scaling laws.}
\end{figure*}

\noindent Hall scaling is the most stable result in the database: \textit{b} $=$ 0$.$991 for all 203 records,\textit{ }\textit{R}\textsuperscript{2} $=$ 0$.$993, and the log-residual scatter factor is 1.16. Resampling the individual records gives a 95\% interval of 0.979–1.004, a narrow range that indicates the result does not depend on a handful of records. Gridded-ion thrust is super-linear in the compiled sample, with \textit{b} $=$ 1$.$199 for 63 steady electric records,\textit{ R}\textsuperscript{2} $=$ 0$.$945 and a scatter factor of 1.75. FEEP is less constrained: \textit{b} $=$ 0$.$805,\textit{ }\textit{R}\textsuperscript{2} $=$ 0$.$695 and a scatter factor of 2.69, and its resampling interval extends from 0.529 to 1.005 – wide enough that both a flatter and a steeper true trend remain plausible. A high coefficient of determination over several decades of power can therefore coexist with poor predictive precision.

MPD gives the clearest example of a result that depends on which sources are trusted. The all-record exponent is 1.078. After keeping only the 11 verified records, it falls to 0.879. The individual-record resampling intervals run from 0.980 to 1.183 for all records and from 0.749 to 0.958 for the verified subset; the document-level check in Table 4 gives closely matching ranges, 0.958–1.193 and 0.743–0.957. Both checks move the result from “consistent with linear” to “below linear” once only verified sources are kept. This does not establish a universal MPD law; it shows that a conclusion drawn from this compilation can depend on which sources are trusted.

The gridded-ion estimate remains above one after the evidence filter, while Hall changes negligibly. The verified FEEP estimate moves further below one, but its wide interval and scatter show that the small, source-dependent sample is inadequate for confident prediction. The more defensible reporting unit is therefore the exponent, its interval, the scatter factor and the sample size together, not\textit{ R}\textsuperscript{2} alone.

\begin{figure*}[t]
\centering
\includegraphics[width=\textwidth]{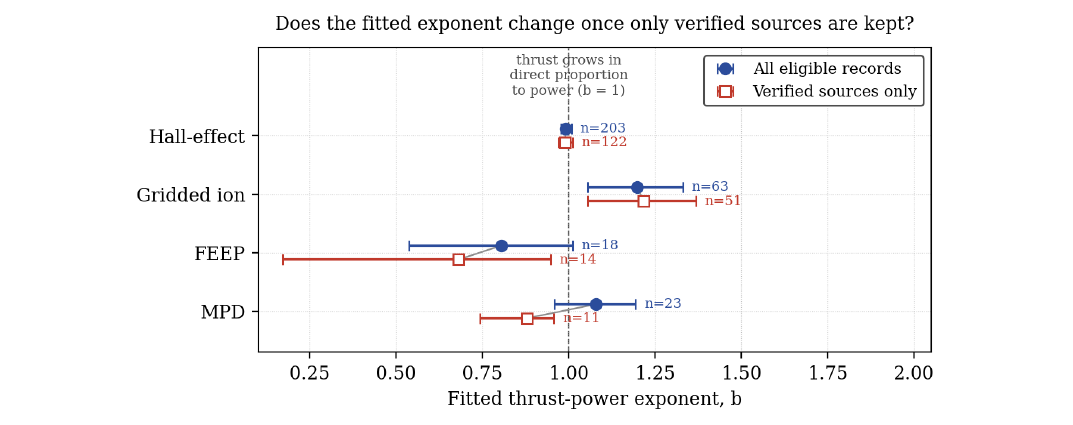}
\caption{Sensitivity of the fitted thrust-power exponent to source grade, for each family with at least eight eligible records. A filled circle marks the exponent fitted from all eligible records. An open square marks the exponent fitted from verified sources only. Each bar is the 95\% interval obtained by resampling whole cited documents 3,000 times, so that several records from the same publication cannot be counted as independent evidence. This is the same check reported in Table 4. A thruster type whose two markers stay close together, with overlapping bars, is one where the result does not depend on how strictly the evidence is filtered.}
\end{figure*}
\begin{table*}[t]
\caption{Source-cluster sensitivity of selected thrust–power fits. Citation texts, rather than individual rows, are resampled. Verified-only fits are reported where at least eight records remain.}
\small\centering
\begin{tabular}{@{}lrrrrrr@{}}
\toprule
 & \multicolumn{3}{c}{All eligible records} & \multicolumn{3}{c}{Verified records only}\\
\cmidrule(lr){2-4}\cmidrule(l){5-7}
Family & $n$ & $b$ & Cluster 95\% interval & $n$ & $b$ & Cluster 95\% interval\\
\midrule
Hall & 203 & 0.991 & 0.976–1.008 & 122 & 0.989 & 0.971–1.011\\
Gridded ion & 63 & 1.199 & 1.055–1.331 & 51 & 1.217 & 1.055–1.368\\
FEEP & 18 & 0.805 & 0.538–1.011 & 14 & 0.680 & 0.171–0.947\\
MPD & 23 & 1.078 & 0.958–1.193 & 11 & 0.879 & 0.743–0.957\\
\bottomrule
\end{tabular}
\end{table*}

Figure 3 plots the same comparison shown in Table 4: for each family, a filled circle marks the exponent\textit{ }\textit{b} fitted from all eligible records and an open square marks the exponent fitted from verified records only, joined by a line so the reader can see at a glance whether the two estimates move together or apart. The bar through each marker is the corresponding 95\% interval from resampling whole cited documents rather than individual rows (Section 3.4), which is the more cautious of the two checks because it does not let several records from the same publication count as independent evidence. A family whose filled and open markers sit close together, with overlapping bars, is one where the fitted exponent does not depend much on how strictly the evidence is filtered; a family whose markers move apart, or whose bars barely overlap, is one where the conclusion depends on which sources are trusted. On this figure, Hall is the stable case: its two markers and their bars overlap almost completely. MPD and FEEP are the sensitive cases: their open-square (verified-only) estimates sit visibly below the filled-circle (all-records) estimates, matching the drop from 1.078 to 0.879 for MPD described above. The central point of the figure is therefore not any single number, but the movement itself: provenance is not documentation added after an analysis, it can change the analysis.

\subsection{Efficiency distributions}

\begin{figure*}[t]
\centering
\includegraphics[width=\textwidth]{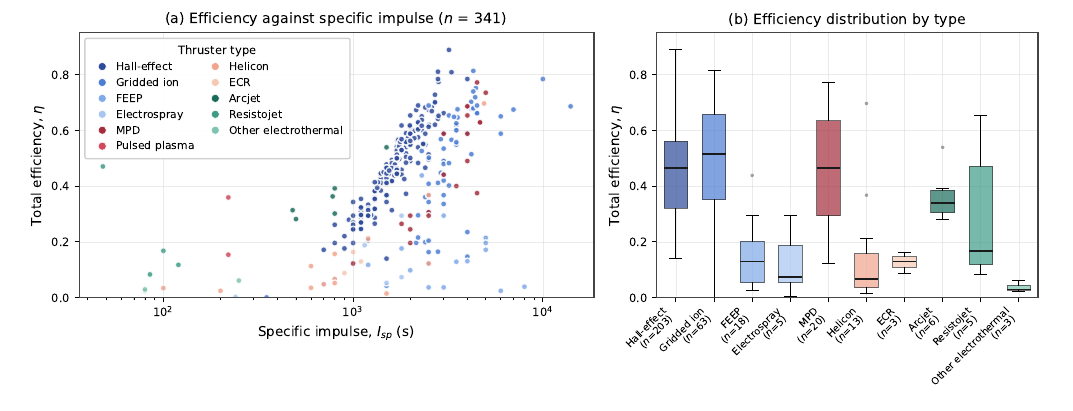}
\caption{Total efficiency across the steady electric subset. Panel (a) plots stored efficiency against specific impulse; panel (b) summarises family distributions. The plotted efficiencies are reconciled database values, not a set of independent measurements with common uncertainty.}
\end{figure*}

\noindent Figure 4 shows wide within-family spreads and a broad association between efficiency and specific impulse. Both features require caution. Test conditions, power boundaries and efficiency definitions are not stored systematically, and the reconciliation procedure makes efficiency algebraically dependent on\textit{ }\textit{P},\textit{ }\textit{T} and\textit{ }\textit{I}\textsubscript{sp} for most eligible rows. Efficiency should not be entered as an additional independent predictor in a model using those same variables unless its measurement lineage is known. The family distributions are nevertheless useful for screening implausible records and locating sparse regions. They show why a single family-average efficiency is inadequate for system sizing. A future release should separate reported total, anode, electrical and thrust efficiencies and retain the inputs and uncertainty for every derived value.

\subsection{Technology Readiness and documented flight of systems}

\noindent The exact TRL column contains six values: 16 records at TRL 3, 130 at TRL 4, 14 at TRL 5, 19 at TRL 7, 97 at TRL 8 and 114 at TRL 9. No released row is assigned TRL 1, 2 or 6. Grouped results therefore comprise 16 concept, 144 development and 230 TRL 7–9 records. Only 134 records contain a first-flight year. Of the 230 rows at TRL 7–9, 133 are dated and 97 are not; all 97 undated rows are assigned TRL 8, reflecting a qualification label rather than a recorded mission year. Conversely, CAT-CubeSat contains a 2018 flight year but is assigned TRL 4. These cases illustrate why the rule, basis string and mission evidence must be inspected together.

\begin{figure*}[t]
\centering
\includegraphics[width=\textwidth]{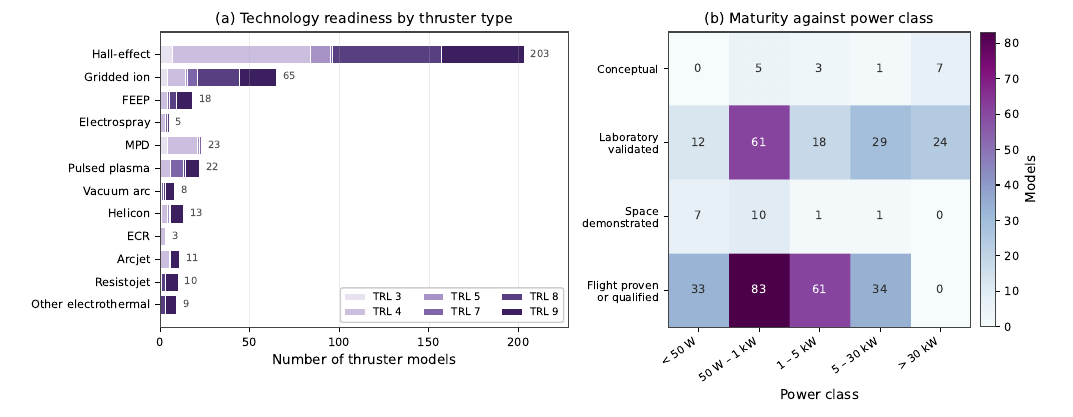}
\caption{Distribution of database readiness. Panel (a) gives exact TRL counts by family; panel (b) cross-tabulates the pipeline’s maturity track and power class. Labels reproduce the database rubric and should not be treated as an independent qualification assessment.}
\end{figure*}

Figure 5 shows this classification in two panels, both built directly from the counts above. Panel (a) is a bar chart of the exact TRL counts (3, 4, 5, 7, 8 and 9) split out by thruster family, so a reader can see at a glance which families are still mostly at concept or development stage and which already have many TRL 7–9 records. Panel (b) is a cross-tabulation: it sorts every record by its grouped maturity track (concept, development or TRL 7–9) and by its power class, and reports how many records fall in each combination. Reading down a column shows how a given power class is distributed across maturity, and reading across a row shows which power classes dominate a given maturity track. Both panels reuse the same rule-based TRL and power-class labels defined earlier in this section, so the figure describes how the database itself classifies each record; it does not certify the real-world readiness of any hardware.

The database assigns 31 records to its “very-high” power class: five are exactly 30 kW and 26 are above 30 kW. This exposes a boundary-definition issue because the stored label says “$>$ 30 kW” although the implemented bin includes 30 kW. None of the 31 records is assigned TRL 7–9; the largest nominal power in that readiness group is 26 kW. These are boundaries of the compiled sample, not proof that no such hardware exists. The class label should be corrected to “30 kW and above” in the next release, and the underlying sources and system boundary should be inspected before this region is used in a mission study.

\subsection{How source selection changes the maturity comparison}

\begin{table}[t]
\caption{Source-use class by database readiness band.}
\footnotesize\centering\setlength{\tabcolsep}{4pt}
\begin{tabular}{@{}lrrrr@{}}
\toprule
Evidence class & Concept & Development & TRL 7–9 & Total\\
\midrule
Verified & 7 & 96 & 143 & 246\\
Manufacturer & 0 & 8 & 41 & 49\\
Compiled & 8 & 30 & 37 & 75\\
Unviewed & 1 & 10 & 9 & 20\\
\midrule
Total & 16 & 144 & 230 & 390\\
\bottomrule
\end{tabular}
\end{table}
\begin{table*}[t]
\caption{Descriptive concept-to-higher-readiness comparison and its sensitivity to source filtering. Ratios are medians for recorded TRL 7–9 divided by medians for TRL 3; they do not represent within-device development trajectories.}
\small\centering
\begin{tabular}{@{}llrrrrrrrr@{}}
\toprule
 & & \multicolumn{2}{c}{Sample size} & \multicolumn{3}{c}{Nominal power (W)} & \multicolumn{3}{c}{Specific impulse (s)}\\
\cmidrule(lr){3-4}\cmidrule(lr){5-7}\cmidrule(l){8-10}
Family & Evidence & TRL 3 & TRL 7–9 & TRL 3 & TRL 7–9 & Ratio & TRL 3 & TRL 7–9 & Ratio\\
\midrule
Hall & All records & 7 & 108 & 2,500 & 1,425 & 0.570 & 1,900 & 1,700 & 0.895\\
Hall & Verified & 5 & 61 & 100,000 & 1,500 & 0.015 & 3,000 & 1,750 & 0.583\\
Gridded ion & All records & 4 & 50 & 550 & 775 & 1.409 & 3,750 & 3,000 & 0.800\\
Gridded ion & Verified & 2 & 41 & 13,800 & 750 & 0.054 & 5,250 & 3,000 & 0.571\\
\bottomrule
\end{tabular}
\end{table*}

\noindent Maturity comparisons are especially sensitive to which sources are kept, because the database is neither a longitudinal study nor a random sample. Table 6 compares the median nominal power and specific impulse of concept records (TRL 3) with the recorded TRL 7–9 group for Hall and gridded-ion thrusters, then repeats the comparison after keeping only verified sources. The ratios change sharply because a single high-power concept record can dominate a small median. The Hall verified-source power ratio is only 0.015 because its small concept subset contains a 100 kW-class programme, not because power normally falls by that factor as a thruster matures. The two verified gridded-ion concepts are similarly too few for a population statement. These results rule out a causal reading of the maturity comparison and reinforce the need to report evidence strata, sample sizes and documented mission events separately. Table 5 confirms that evidence quality and readiness are independent of each other. A verified source may describe a low-readiness laboratory experiment, while a manufacturer source may describe qualified hardware. Collapsing both dimensions into a single score would hide rather than simplify the evidence.

\subsection{Mass, historical coverage and review flags}

\begin{figure*}[t]
\centering
\includegraphics[width=\textwidth]{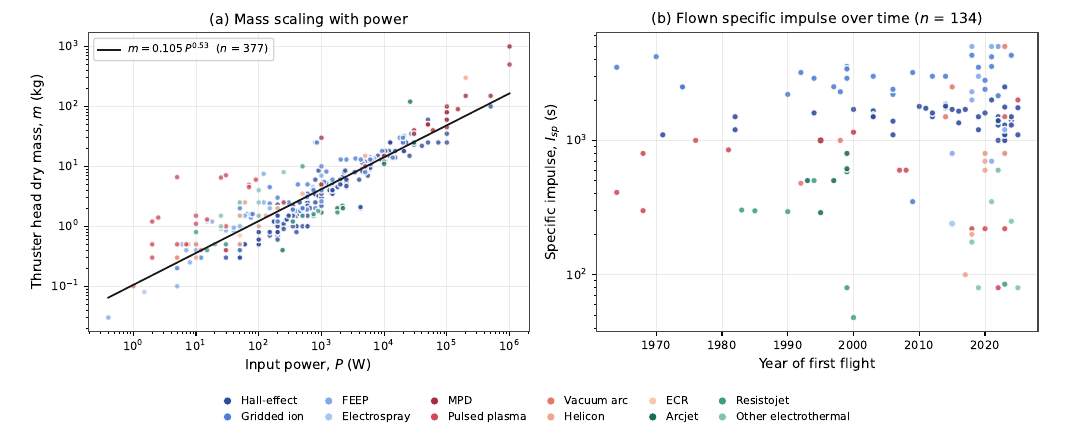}
\caption{Mass and historical coverage in the supplied database. Panel (a) shows recorded dry mass against power and the pooled descriptive fit; panel (b) shows specific impulse for the 134 records with a first-flight year. Family, mass boundary and missing-date effects prevent causal interpretation.}
\end{figure*}

\noindent Panel (a) of Figure 6 plots the recorded dry mass\textit{ m} against nominal power\textit{ }\textit{P} for every record that has both values, one point per thruster, on logarithmic axes. The fitted line,\textit{ m} $=$ 0$.$105\textit{P}\textsuperscript{0}$^{.}$\textsuperscript{53} for 377 records, is calculated the same way as the thrust–power fit in Section 3.4 and Figure 2: take the logarithm of mass and the logarithm of power for every record, fit a straight line to log \textit{m} against log \textit{P} by ordinary least squares, and convert the fitted intercept and slope back into the coefficient 0$.$105 and the exponent 0$.$53. An exponent below one means that, across the whole compiled sample, mass grows more slowly than power does. This is a description of the compiled sample only. The schematic does not consistently distinguish thruster-head mass from a full propulsion assembly, and both family and programme scale co-vary with power, so the fitted exponent must not be used as a universal component-sizing law without redefining the mass boundary and validating the fit within individual families.

Panel (b) of Figure 6 shows specific impulse plotted against first-flight year for the 134 records that have a recorded year, so it is a historical view of the same database rather than a second mass fit. The historical data span 1964–2025 but omit 256 records with no recorded year. The apparent broadening of the spread after 1990 may reflect real programme history, improved documentation in more recent sources, or simple collection bias toward better-documented modern hardware, and the figure cannot distinguish between these explanations on its own. A defensible historical analysis would need mission identity, first launch, first successful operation and routine-use dates as separate fields rather than the single first-flight year used here.

Sixty-eight rows belong to 28 exact repeated power–thrust–specific-impulse groups. These may be valid variants, aliases or copied representative values, so no record is deleted solely for matching another. One family assignment, VELARC, remains flagged for review because its ammonia propellant and cited very-low-power arcjet description conflict with the stored vacuum-arc label. The release preserves the issue rather than applying an unlogged correction.

The dataset also contains 22 pulsed-plasma and eight vacuum-arc rows, but it lacks impulse bit, mass bit, pulse energy, repetition rate, pulse duration, peak power and duty cycle. For repeated pulses, average thrust is\textit{ T}\textsubscript{avg}$=$\textit{ I}\textsubscript{bit}\textit{ f}, which cannot be checked without impulse bit and repetition frequency (Molina-Cabrera et al. 2011). No row can therefore be identified safely as one-shot or assigned a unique duty cycle from the current fields.

\subsection{Gaussian-process analysis of efficiency residuals}

\begin{figure*}[t]
\centering
\includegraphics[width=0.93\textwidth]{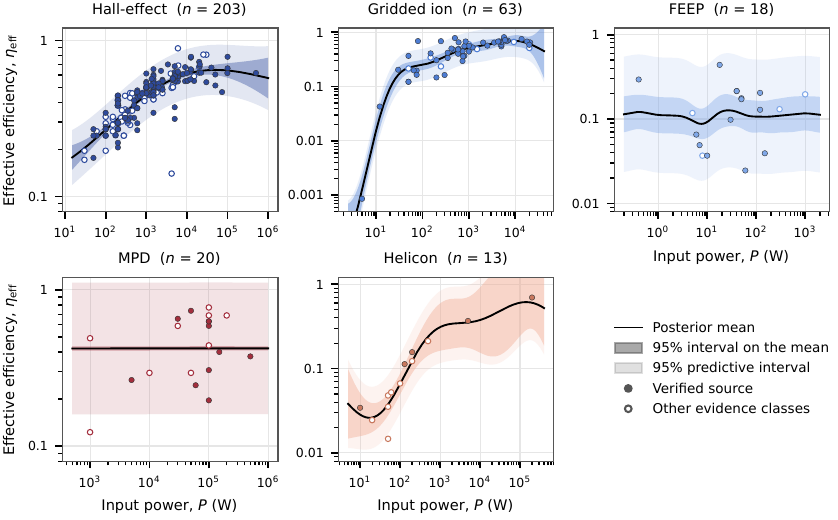}
\caption{Gaussian-process posterior of effective efficiency against input power (model A) for the steady electric families with at least eight eligible records. The dark band is the 95\% interval on the posterior mean and the light band is the 95\% predictive interval, which includes the fitted noise. Curves extend beyond the observed power range to show the growth of posterior variance.}
\label{fig:gp}
\end{figure*}

\begin{table*}[t]
\caption{Held-out performance over five contiguous power regions. Error factors are multiplicative root-mean-square errors on thrust. Coverage is the percentage of held-out records inside the 95\% predictive interval. Model A uses power; model B adds specific impulse; model C adds propellant class, maturity track and evidence class.}
\label{tab:gp}
\centering\small
\begin{tabular}{@{}lrccccccccc@{}}
\toprule
 & & \multicolumn{5}{c}{Error factor} & \multicolumn{3}{c}{Coverage (\%)}\\
\cmidrule(lr){3-7}\cmidrule(l){8-10}
Family & $n$ & Constant $\eta$ & Equation 6 & A & B & C & A & B & C\\
\midrule
Hall-effect & 203 & 1.51 & 1.17 & 1.31 & 1.18 & 1.18 & 91.1 & 93.6 & 93.1\\
Gridded ion & 63 & 2.88 & 2.15 & 2.54 & 2.68 & 2.30 & 85.7 & 81.0 & 88.9\\
FEEP & 18 & 2.29 & 2.71 & 2.34 & 2.34 & 2.03 & 94.4 & 94.4 & 66.7\\
MPD & 20 & 1.76 & 1.46 & 1.76 & 1.43 & 1.53 & 95.0 & 90.0 & 65.0\\
Helicon & 13 & 3.81 & 3.54 & 2.83 & 2.78 & 2.77 & 76.9 & 76.9 & 61.5\\
\midrule
\multicolumn{10}{@{}l}{Verified-source subset}\\
Hall-effect & 122 & 1.48 & 1.20 & 1.26 & 1.15 & 1.14 & 92.6 & 94.3 & 90.2\\
Gridded ion & 51 & 3.22 & 2.29 & 2.82 & 2.96 & 2.60 & 84.3 & 80.4 & 80.4\\
FEEP & 14 & 2.38 & 2.68 & 2.40 & 2.40 & 1.88 & 92.9 & 92.9 & 64.3\\
MPD & 11 & 1.56 & 1.27 & 1.56 & 1.32 & 1.59 & 100.0 & 45.5 & 45.5\\
\bottomrule
\end{tabular}
\end{table*}

The analysis described in Section 2.5 was applied to the 345 steady electric records. Four records with an effective efficiency above one (AAC-EPSS, MagBeam, MagBeam-Concept and NASA-1MW-MPD) were returned for source review, all of which carry the irreconcilable label. The remaining 341 records have the efficiency values shown in Figure 4. Separate models were fitted to the five families with at least eight eligible records, using a squared-exponential kernel with automatic relevance determination and a noise term on $\log_{10}\eta_{\mathrm{eff}}$. Three input sets were compared. Model A uses input power, model B adds specific impulse and model C adds propellant class, maturity track and evidence class. Subtype was evaluated separately because only 59 of the 341 records carry a subtype label.

Table~\ref{tab:gp} lists held-out performance for five contiguous power regions, with records at the same power kept in the same region. Errors are expressed as multiplicative factors on thrust and are compared with two references: a constant family efficiency and the power-law form of Equation 6. Specific impulse gives the largest improvement for Hall-effect records, where the error factor falls from 1.31 for model A to 1.18 for model B, close to the 1.17 obtained from Equation 6, with 93.6\% of held-out records inside the 95\% predictive interval. The same step reduces the MPD factor from 1.76 to 1.43 with 90.0\% coverage. Larger errors remain for gridded-ion and helicon records, at 2.68 and 2.78 for model B with coverage of 81.0\% and 76.9\%. Model C lowers the gridded-ion factor to 2.30 and raises its coverage to 88.9\%. In the smaller FEEP and MPD samples, the same inputs reduce coverage to 66.7\% and 65.0\%, which indicates overconfident intervals when categorical inputs are added to few records.

The posterior of model A is shown in Figure~\ref{fig:gp}. The Hall-effect mean rises from an effective efficiency of about 0.20 at 30 W to 0.61 at 10 kW and then flattens, which is consistent with the expected reduction of wall and ionisation losses with thruster scale. The posterior variance widens beyond the observed power range and across sparsely sampled intervals, such as the helicon range between 500 W and 200 kW. The fitted signal amplitude is negligible for MPD records, so power alone explains little of the efficiency variation in that family within this sample. The steep gridded-ion decline below 30 W is controlled by a small number of low-power records.

Repeating the analysis with verified sources gives a consistent Hall-effect result, with an error factor of 1.15 and coverage of 94.3\% for model B. The gridded-ion factor increases to 2.96 with coverage of 80.4\%, and the FEEP factor is 2.40 with coverage of 92.9\%. The verified MPD factor falls to 1.32, although coverage falls to 45.5\% with 11 records, so this interval is unsuitable for use. Holding out each labelled subtype in turn with model B gives factors of 1.11 for Hall-effect (17 records in six subtypes), 1.60 for gridded ion (10 records in three subtypes), 3.25 for FEEP (seven records in five subtypes) and 1.21 for MPD (five records in two subtypes). A two-stage heteroscedastic model gave a slightly poorer held-out log density for Hall-effect records and a small improvement for gridded-ion records, with coverage of 82.5\%. Input-dependent noise is therefore retained as a candidate for gridded-ion thrusters only. These results support model B for preliminary thrust screening of Hall-effect records within the observed power range. Calibrated use for the other families requires additional physical inputs and operating-point data.

\subsection{Mission-level application of the database}

\begin{figure*}[t]
\centering
\includegraphics[width=\textwidth]{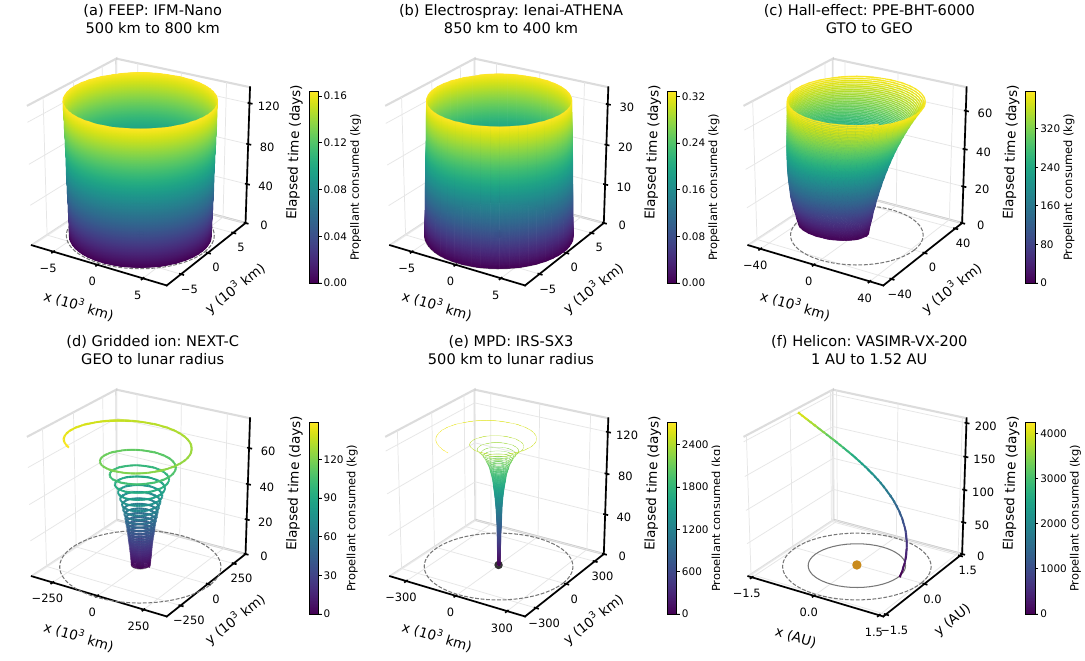}
\caption{Low-thrust transfers propagated from six database records. The horizontal axes give position in the orbital plane and the vertical axis gives elapsed time. Colour gives propellant consumed. Grey curves on the base plane mark the initial circular orbit (solid) and the target radius (dashed). Inputs and results are listed in Table~\ref{tab:transfers}.}
\label{fig:transfers}
\end{figure*}

\begin{table*}[t]
\caption{Database inputs, assumed mission inputs and propagated results for the transfers in Figure~\ref{fig:transfers}. $P$, $T$ and $I_{\mathrm{sp}}$ refer to one thruster, $n$ is the cluster size, $m_0$ is the initial mass, $\Delta V$ is the integrated velocity increment and $e$ is the eccentricity on arrival; Prop.\ denotes propellant consumed. VASIMR VX-200 is classed as helicon in the database.}
\label{tab:transfers}
\centering\footnotesize
\setlength{\tabcolsep}{3pt}
\resizebox{\textwidth}{!}{%
\begin{tabular}{@{}llrrrrrrrrrrr@{}}
\toprule
Record & Transfer & $P$ (W) & $T$ (mN) & $I_{\mathrm{sp}}$ (s) & TRL & $n$ & $m_0$ (kg) & $nP$ (kW) & $\Delta V$ (km/s) & Time (d) & Prop. (kg) & $e$\\
\midrule
(a) IFM-Nano & 500 to 800 km & 40 & 0.35 & 5,000 & 9 & 2 & 50 & 0.08 & 0.16 & 132.7 & 0.16 & 0.000\\
(b) Ienai ATHENA & 850 to 400 km & 15 & 0.50 & 1,800 & 4 & 4 & 24 & 0.06 & 0.24 & 33.5 & 0.33 & 0.000\\
(c) PPE BHT-6000 & GTO to GEO & 6,000 & 350 & 2,200 & 8 & 4 & 4,000 & 24 & 2.25 & 70.7 & 396 & 0.002\\
(d) NEXT-C & GEO to lunar radius & 6,900 & 236 & 4,190 & 9 & 4 & 3,000 & 27.6 & 2.08 & 74.7 & 148 & 0.224\\
(e) IRS SX3 & 500 km to lunar radius & 100,000 & 2,750 & 4,665 & 4 & 4 & 20,000 & 400 & 6.66 & 130.4 & 2,710 & 0.394\\
(f) VASIMR VX-200 & 1 to 1.52 AU & 200,000 & 5,800 & 4,900 & 5 & 2 & 30,000 & 400 & 7.31 & 202.9 & 4,233 & 0.404\\
\bottomrule
\end{tabular}}
\end{table*}

The database supplies the thrust, specific impulse and input power required to propagate a low-thrust transfer from any record with matched fields. Six records were selected to demonstrate this use across the power range of the release: IFM-Nano (FEEP, 40 W), Ienai ATHENA (electrospray, 15 W), PPE BHT-6000 (Hall-effect, 6 kW), NEXT-C (gridded ion, 6.9 kW), IRS SX3 (MPD, 100 kW) and VASIMR VX-200 (200 kW). Each record is scaled by the number of thrusters in a cluster, $n$, which gives a total thrust $nT$ and a required electrical power $nP$. The spacecraft motion is integrated in the orbital plane from
\begin{equation}
\ddot{\mathbf{r}}=-\frac{\mu}{r^{3}}\mathbf{r}+\frac{nT}{m}\,\hat{\mathbf{u}},\qquad \dot{m}=-\frac{nT}{g_0I_{\mathrm{sp}}},
\end{equation}
where $\mu$ is the gravitational parameter of the central body and $\hat{\mathbf{u}}$ is the thrust direction. A fourth-order Runge--Kutta scheme is used with 800 steps per local circular period. Thrust is tangential in the spiral transfers and retrograde in the orbit-lowering case. The transfer from geostationary transfer orbit to geostationary orbit uses locally optimal steering on semi-major axis and eccentricity. Spiral transfers end when the spacecraft reaches the target radius, and capture and circularisation are excluded from the calculation.

The resulting trajectories are presented in Figure~\ref{fig:transfers}, where the horizontal axes give the position in the orbital plane, the vertical axis gives elapsed time and the colour scale gives propellant consumed. The inputs and outputs of each case are listed in Table~\ref{tab:transfers}. Transfer durations range from 34 days for the electrospray case to 203 days for the VASIMR case, and propellant mass ranges from 0.16 kg to 4,233 kg. The trajectory geometry follows the ratio of thrust acceleration to local gravity. The FEEP and electrospray cases complete several hundred to almost two thousand revolutions at nearly constant radius, whereas the MPD case widens from a dense inner spiral to a few large revolutions near the lunar radius. The VASIMR case completes half a revolution around the Sun.

Elapsed time scales with $1/n$ at a fixed velocity increment and propellant mass remains constant, so cluster size exchanges transfer time against bus power. The velocity increments of the FEEP, electrospray, gridded-ion and MPD cases agree with the coplanar Edelbaum value (Edelbaum 1961) to within 1.3\%. The VASIMR case exceeds this value by 29\% and arrives with an eccentricity of 0.40, which is consistent with its higher ratio of acceleration to local gravity. Doubling the number of steps per period changes each transfer time by less than 0.4\%.

Cluster size and initial mass are assumed mission inputs. Power is held constant, and eclipses, the decline of solar power with distance, perturbations, plane changes, power-system mass and thruster lifetime are excluded. The Ienai ATHENA, IRS SX3 and VASIMR VX-200 records lie below TRL 7, so panels (b), (e) and (f) describe potential capability at their current stage of development. These cases demonstrate the workflow and provide preliminary estimates. The same procedure applies to every record with matched power, thrust and specific impulse, which allows a mission designer to compare thruster type, operating point and cluster size against transfer time, propellant mass and bus power from one traceable dataset.

\section{Discussion}

\subsection{A focused and defensible contribution}

\noindent The primary contribution is a reproducible evidence chain across a heterogeneous technology space. It connects a preserved source compilation to explicit exclusion and collapse rules, a citation-gated release, source-use classes, physical checks, analysis code and sensitivity results. It does not claim that prior databases are absent, that every value has been independently re-measured, or that a new law of thruster physics has been discovered.

This framing matters because the strongest result is methodological as well as empirical. The MPD scaling change demonstrates that provenance can alter an engineering conclusion. The Hall result shows the complementary case: a relation that remains stable under both evidence filtering and citation-cluster resampling. Reporting both outcomes is more informative than presenting a single best-fit line without its evidence dependence.

The paper also clarifies the database’s place within the wider PhD. The release is the evidence layer for a future unified systems-engineering toolkit. It can define candidate regions and expose where mission, power and propulsion assumptions require stronger data. It is not itself an operational digital twin, and no telemetry-driven update loop or independent experimental calibration is claimed here. Keeping that boundary explicit protects the minimum viable research contribution from being diluted by later architectural work.

\subsection{Evaluation and use cases of the database}

\noindent The current database can support descriptive technology maps that involve checking source quality and completeness, early screening by recorded power, thrust, specific impulse and mass, identification of influential or weakly evidenced records, and cautious within-family regressions with evidence sensitivity. It is also suitable for prioritising which sources and families should be re-extracted first.

It cannot support hardware qualification, procurement, causal claims about maturity, or autonomous selection. Nor can it replace subsystem engineering. Lifetime, total impulse, throughput, flow rate, voltage, current, plume divergence, facility pressure, contamination, thermal interfaces, electromagnetic compatibility, pointing, duty cycle, tankage, power-processing mass, availability and cost can all change a technology decision. The database can identify candidates; however, it cannot finalise choosing a mission design.

Scaling exponents should therefore be described as database regressions. Device-held-out validation and independent external data are decisive before any regression is embedded in a predictive sizing tool. Extrapolation beyond a family’s observed power range is especially weak, as thruster-type-specific machine-learning studies have also shown (Marianacci and Mazouffre 2024).

\subsection{Power-bus coupling and mission context}

\noindent Equation 5 shows why thruster selection cannot be separated from spacecraft power. At a specified\textit{ I}\textsubscript{sp} and total efficiency, greater thrust requires proportionally greater electrical input; the spacecraft must also provide conversion, distribution and thermal rejection with appropriate duty cycle and margin. The database can expose this coupling, but its nominal thruster power is not a complete spacecraft-bus requirement.

VASIMR illustrates the distinction between a ground thruster result and an integrated flight capability. The released VX-200 row records a 200 kW nominal point and rule-assigned TRL 5. Ad Astra reported that the related VX-200SS prototype operated continuously for 88 hours at 80 kW in 2021 and described the engine as TRL 5 (Ad Astra Rocket Company 2021). This is relevant endurance evidence, but it does not demonstrate a 200 kW spacecraft power, processing and thermal architecture. The example is therefore used to motivate subsystem budgeting, not to label the programme a failure.

A future sizing toolkit should apply mission duration, available power, duty cycle, total impulse and spacecraft constraints before ranking candidates. It may compare clustered units or architectures containing more than one propulsion family where the mission and control concept justify them. The present database supplies a common evidence layer for that comparison; it does not yet validate a hybrid architecture. It also contains no mission-population denominator or causal outcome variable, so it cannot support a quantified claim that weak power coupling produces a higher failure rate.

\subsection{Limitations and improvements}

\noindent The study has five principal limitations. First, one citation supports a row rather than each field, so citation completeness exceeds claim-level traceability. Second, one representative row cannot preserve a throttle map or guarantee matched operating conditions. Third, source grade records document handling, not measurement uncertainty. Fourth, family, power, date and maturity are confounded in a non-random literature sample. Fifth, physical definitions such as electrical power boundary, efficiency basis and dry-mass boundary are incomplete.

Independent checking should accompany the schema change. A stratified sample should be extracted twice by different reviewers, with field-level agreement and disagreements reported. Device-held-out validation should then test predictive use, and external facility or flight data should estimate model discrepancy rather than being used only for visual confirmation. A future telemetry update path can be defined architecturally, but it should not be described as a demonstrated digital twin until independent data close the loop.

The next research step is therefore stronger evidence, not additional visual complexity: exact source locators, matched conditions, uncertainty, independent validation and a defensible update rule. These additions would allow the database to progress from a reproducible cross-family inventory to a reliable evidence layer for mission–power–propulsion sizing.

\section{Conclusion}

\noindent This paper is an overview and analysis of the electric propulsion database. This study developed and evaluated a reproducible database of 390 named EP systems across 12 thruster types. The processing history shows how 433 legacy performance records were consolidated into 306 named models, expanded with documented candidates and reduced to the final release after six uncited candidates were withdrawn. Every released row has a full citation and 80.3\% include a direct link. These controls provide a traceable basis for further analysis.

The results show that the database is most useful for comparative mapping, evidence review and early engineering screening. The thrust–specific-impulse map identifies the recorded technology regions, while the thrust-to-power relation explains why EP selection must be linked to the spacecraft power budget and mission. The efficiency and mass results are descriptive rather than independent physical validation because some variables were reconciled algebraically and the system boundaries are incomplete. The absence of TRL 7–9 records among the 31 systems at or above 30 kW indicates limited high-power maturity within this sample; it does not establish a universal technology limit.

Source quality has a measurable effect on the scaling analysis. The Hall thrust–power exponent remains stable after source filtering, but the MPD exponent falls from 1.078 for all records to 0.879 for verified records. The interpretation therefore changes from approximately linear to sub-linear within the verified sample. The gridded-ion trend remains above linear in the database, while the FEEP result remains uncertain because its sample is small and variable. These findings show that provenance must be treated as part of the analysis, not only as supporting documentation.

The present release is suitable for descriptive comparison and carefully qualified candidate screening, but not for hardware qualification, procurement or automatic technology selection. The next version should add field-level source locations, condition-resolved throttle data, uncertainty, pulse parameters, clear system boundaries and independent validation. With these additions, the database can provide a stronger foundation for a mission-constrained sizing toolkit. Telemetry-based digital-twin development remains outside the scope of this paper.

\section*{Data availability}
The database is archived on Zenodo (Space Systems Research Group 2026) at \url{https://doi.org/10.5281/zenodo.23948174}. The companion interactive application, EP Mission Designer, is available at \url{https://ep-integrated-mission-design-tool.streamlit.app/}.

\section*{References}
\small
\begin{list}{}{\setlength{\leftmargin}{12pt}\setlength{\itemindent}{-12pt}\setlength{\itemsep}{2pt}\setlength{\parsep}{0pt}\setlength{\topsep}{0pt}}\raggedright
\item Ad Astra Rocket Company. VX-200SS VASIMR prototype completes 88-hour endurance test at 80 kw. Technical report, Ad Astra Rocket Company, October 2021. URL \url{https://www.adastrarocket.com/wp-content/uploads/2021/10/Press-Release-October-13-2021-final.pdf}. Manufacturer press release; reports TRL 5.
\item Michelle Barker, Neil P. Chue Hong, Daniel S. Katz, et al. Introducing the FAIR principles for research software. \textit{Scientific Data}, 9:622, 2022. doi: 10.1038/s41597-022-01710-x.
\item Kathe Dannenmayer and Stephane Mazouffre. Elementary scaling laws for sizing up and down hall effect thrusters: Impact of simplifying assumptions. In \textit{Proceedings of the 31st International Electric Propulsion Conference}, Ann Arbor, MI, 2009. URL \url{https://electricrocket.org/IEPC/IEPC-2009-077.pdf}. Paper IEPC-2009-077.
\item Theodore N. Edelbaum. Propulsion requirements for controllable satellites. \textit{ARS Journal}, 31(8):1079--1089, 1961. doi: 10.2514/8.5723.
\item Guido Giammarinaro, Francesco Marconcini, Giulia Becatti, Manuel M. Saravia, Mariano Andrenucci, and Fabrizio Paganucci. A scaling methodology for high-power magnetically shielded hall thrusters. \textit{Journal of Electric Propulsion}, 2:17, 2023. doi: 10.1007/s44205-023-00049-8.
\item Dan M. Goebel and Ira Katz. \textit{Fundamentals of Electric Propulsion: Ion and Hall Thrusters}. John Wiley \& Sons and Jet Propulsion Laboratory, Hoboken, NJ, 2008. URL \url{https://descanso.jpl.nasa.gov/SciTechBook/series1/Goebel__cmprsd_opt.pdf}.
\item David Krejci and Paulo Lozano. Space propulsion technology for small spacecraft. \textit{Proceedings of the IEEE}, 106(3):362–378, 2018. doi: 10.1109/JPROC.2017.2778747.
\item Igor Levchenko, Kateryna Bazaka, Yongjie Ding, et al. Space micropropulsion systems for CubeSats and small satellites: From proximate targets to furthermost frontiers. \textit{Applied Physics Reviews}, 5(1):011104, 2018. doi: 10.1063/1.5007734.
\item Alfredo Marianacci and Stephane Mazouffre. Supervised machine learning-based hall thruster scaling. \textit{Journal of Electric Propulsion}, 3:14, 2024. doi: 10.1007/s44205-024-00077-y.
\item Silvia Masillo, Andrea Lucca Fabris, and Paolo Bianco. A semi-empirical scaling model for low-power hall effect thrusters and its application to the analysis of the halo thruster performance. \textit{Journal of Electric Propulsion}, 4:67, 2025. doi: 10.1007/s44205-025-00165-7.
\item Pedro Molina-Cabrera, Georg Herdrich, Matthias Lau, Stefanos Fasoulas, Tony Schoenherr, and Kimiya Komurasaki. Pulsed plasma thrusters: A worldwide review and long yearned classification. In \textit{Proceedings of the 32nd International Electric Propulsion Conference}, Wiesbaden, Germany, 2011. URL \url{https://electricrocket.org/IEPC/IEPC-2011-340.pdf}. Paper IEPC-2011-340.
\item NASA Small Spacecraft Systems Virtual Institute. State-of-the-art of small spacecraft technology: In-space propulsion. Technical report, National Aeronautics and Space Administration, 2026. URL \url{https://www.nasa.gov/smallsat-institute/sst-soa/in-space_propulsion/}. Public information current to April 2026; accessed 4 August 2026.
\item National Aeronautics and Space Administration. Technology readiness assessment best practices guide. Technical Report NASA/SP-20205003605, National Aeronautics and Space Administration, 2020. URL \url{https://ntrs.nasa.gov/citations/20205003605}.
\item Dillon O’Reilly, Georg Herdrich, and Darren F. Kavanagh. Electric propulsion methods for small satellites: A review. \textit{Aerospace}, 8(1):22, 2021. doi: 10.3390/aerospace8010022.
\item James E. Polk, Anthony Pancotti, Thomas Haag, Steven King, Mitchell Walker, John Blakely, and Jennifer Ziemer. Recommended practice for thrust measurement in electric propulsion testing. \textit{Journal of Propulsion and Power}, 33(3):539–555, 2017. doi: 10.2514/1.B35564.
\item Princeton University Electric Propulsion and Plasma Dynamics Laboratory. Applied-field MPD thruster database. Public research database, 2026. URL \url{https://alfven.princeton.edu/tools/afmpdt-database/}. Accessed 4 August 2026.
\item Alexander Reissner, Stefan Hummel, and Eric Gengembre. Development status and future of the european electric propulsion plume database. In \textit{Proceedings of the 33rd International Electric Propulsion Conference}, Washington, DC, 2013. URL \url{https://electricrocket.org/IEPC/03hwl93c.pdf}. Paper IEPC-2013-093.
\item Geir Kjetil Sandve, Anton Nekrutenko, James Taylor, and Eivind Hovig. Ten simple rules for reproducible computational research. \textit{PLoS Computational Biology}, 9(10):e1003285, 2013. doi: 10.1371/journal.pcbi.1003285.
\item Space Systems Research Group. Electric propulsion systems database from the Space Systems Research Group. Dataset, Zenodo, 2026. doi: 10.5281/zenodo.23948174.
\item Mark D. Wilkinson, Michel Dumontier, IJsbrand Jan Aalbersberg, et al. The FAIR guiding principles for scientific data management and stewardship. \textit{Scientific Data}, 3:160018, 2016. doi: 10.1038/sdata.2016.18.
\end{list}
\end{document}